\documentclass{article}
\usepackage{times}
\usepackage{amsfonts}
\usepackage{graphicx}
\usepackage[pdfmark]{hyperref}
\begin{document}
\noindent
{\Large RICCI FLOW, QUANTUM MECHANICS AND  GRAVITY}
\vskip1cm
\noindent
{\bf Jos\'e M. Isidro}${}^{a,b,1}$, {\bf J.L.G. Santander}${}^{a,2}$ and {\bf P. Fern\'andez de C\'ordoba}${}^{a,3}$\\
${}^{a}$Grupo de Modelizaci\'on Interdisciplinar, Instituto de Matem\'atica Pura y Aplicada,\\ Universidad Polit\'ecnica de Valencia, Valencia 46022, Spain\\
${}^{b}$Max--Planck--Institut f\"ur Gravitationsphysik, Albert--Einstein--Institut,\\ D--14476 Golm, Germany\\
${}^1${\tt joissan@mat.upv.es}, ${}^2${\tt jlgonzalez@mat.upv.es},\\  ${}^3${\tt pfernandez@mat.upv.es}
\vskip1cm
\noindent
{\bf Abstract} It has been argued that, underlying any given quantum--mechanical model, there exists at least one deterministic system that reproduces, after prequantisation, the given quantum dynamics. For a quantum mechanics with a complex $d$--dimensional Hilbert space, the Lie group $SU(d)$ represents classical canonical transformations on the projective space $\mathbb{CP}^{d-1}$ of quantum states.  Let $R$ stand for the Ricci flow of the manifold $SU(d-1)$ down to one point, and let $P$ denote the projection from the Hopf bundle onto its base $\mathbb{CP}^{d-1}$. Then the underlying deterministic model we propose here is the Lie group $SU(d)$, acted on by the operation $PR$.  Finally we comment on some possible consequences that our model may have on a quantum theory of gravity.

\section{Introduction}\label{intt}

Quantum mechanics as a statistical theory has been argued to {\it emerge}\/ from an underlying deterministic theory \cite{THOOFT1, THOOFT2}. Specifically, for any quantum system there exists at least one deterministic model that reproduces all its dynamics after prequantisation. This existence theorem has been extended  to include cases characterised by sets of commuting beables \cite{ELZE1}; it  has also been complemented with an explicit dynamical theory \cite{ELZE5}. The ultimate goal of all these endeavours is to provide us with a better understanding of quantum gravity (for reviews see, {\it e.g.}, refs. \cite{QGUNO, QGDOS}) and to eventually help us solve some of its puzzles, most notably the information loss paradox \cite{THOOFT3}, the value of the cosmological constant, and the issue of dark energy \cite{PADDY1, PADDY3}. 

Mechanisms have been presented \cite{THOOFT1, THOOFT2, ELZE1, ELZE5} to explain the passage from a deterministic theory to a probabilistic theory. Usually they are based on a dynamical system, the phase--space trajectories of which possess suitably located attractors ({\it e.g.}, at the eigenvalues of the given quantum Hamiltonian, or at certain configurations of the density matrix).  These mechanisms can be thought of as an existence theorem, in that every quantum system (with a finite--dimensional Hilbert space) possesses at least one deterministic system underlying it. 

On the other hand there are plenty of dissipation equations in physics and mathematics, equations implementing the information loss that is characteristic of the passage from classical to quantum. The heat equation, and its close cousin the Schroedinger equation, immediately come to mind. 

Along apparently unrelated lines, but in fact deeply connected with our present quest, we would like to mention the following points. Ricci--flow equations have attracted much attention recently, first and foremost in geometry \cite{PERELMAN, ANDERSON} but also well beyond pure mathematics. Intriguing connections between geometry, Ricci--flow equations, quantum mechanics and gravity have been explored in refs. \cite{CARROLL1, CARROLL2}. In this spirit one should also cite refs. \cite{MATONE1, PADDY2}, where it has been speculated that gravity could possibly have a quantum origin.  Further interesting topics at the interface of gravity and quantum mechanics are dealt with in refs. \cite{ELZE2, ELZE3, ELZE4, CARROLL5, CARROLL6, GP, TIME}; a classical/quantum duality with gravitational side effects has been analysed in refs. \cite{MATONE2, ME, SINGH}. Other issues, in principle outside the realm of quantum gravity and concerning quantum mechanics proper, but in fact intimately related, are interpretational problems such as the measurement problem and the collapse of the wavefunction \cite{MIGUEL}.

In this paper we develop a deterministic model exhibiting dissipation, from which quantum mechanics emerges naturally. Given a quantum mechanics with a complex $d$--dimensional Hilbert space, the Lie group $SU(d)$ represents classical canonical transformations on the projective space $\mathbb{CP}^{d-1}$ of quantum states \cite{GIACHETTA}.  Let $R$ stand for the Ricci flow of the manifold $SU(d-1)$ down to one point, and let $P$ denote the projection from the Hopf bundle onto its base $\mathbb{CP}^{d-1}$. Then the underlying deterministic model we propose here is the Lie group $SU(d)$, acted on by the operation $PR$.

\section{The Ricci flow as a dissipative mechanism}\label{grund}

Given an $n$--dimensional manifold $\mathbb{M}$ endowed with the Riemannian metric $g_{ij}$, the equations governing the (unnormalised) Ricci flow read
\begin{equation}
\frac{\partial g_{ij}}{\partial t}=-2R_{ij}, \qquad i,j=1,\ldots, n,\qquad t\geq 0,
\label{guillaume}
\end{equation}
where $t$ is an evolution parameter (not a coordinate on $\mathbb{M}$), and $R_{ij}$ is the Ricci tensor corresponding to the metric $g_{ij}$. Informally one can say that (Ricci) flat spaces remain unchanged under the flow, while positively curved manifolds contract  and negatively curved manifolds expand under the same flow. We will be interested in the particular case of Einstein manifolds  \cite{BESSE}, where the Ricci tensor and the metric are proportional:
\begin{equation}
R_{ij}=\kappa g_{ij}, 
\label{onestone}
\end{equation}
with $\kappa$ a constant.  Since the metric is positive definite,  the sign of $\kappa$ equals the sign of the Ricci tensor. Relevant examples of positively curved Einstein manifolds are complex projective space $\mathbb{CP}^N$ and the special unitary group $SU(N)$, both of which will play an important role in what follows. Their respective metrics are the Fubini--Study metric \cite{KOBAYASHI} and the Killing--Cartan metric \cite{HELGASON}. 

Under the Ricci flow, the contraction of a whole manifold down to a point can play the role of a dissipative mechanism. One hint that this intuition is correct comes from the following example. Consider a 2--dimensional manifold endowed with the isothermal coordinates $x$ and $y$. The metric reads 
\begin{equation}
{\rm d}s^2={\rm e}^{-f}\left({\rm d}x^2+{\rm d}y^2\right).
\label{isso}
\end{equation}
Allowing the metric to depend also on the evolution parameter $t$, the Ricci flow equation (\ref{guillaume}) becomes
\begin{equation}
\frac{\partial f}{\partial t}=\nabla^2 f.
\label{heiss}
\end{equation}
The above is formally identical to the heat equation, with one important difference, however: the Laplacian $\nabla^2$ is computed with respect to the metric (\ref{isso}), in which it reads
\begin{equation}
\nabla^2 f={\rm e}^{f}\left(\frac{\partial^2 f}{\partial x^2}+\frac{\partial^2f}{\partial y^2}\right).
\label{placa}
\end{equation}
Regardless of the nonlinearity of (\ref{placa}), the fact that the Ricci--flow equation can be recast as a heat equation is a clear hint that a dissipative mechanism is at work.

\section{The underlying deterministic model}\label{fluss}

{}For the rest of this paper we will consider a quantum system with a finite, complex $d$--dimensional Hilbert space of quantum states, that we can identify with $\mathbb{C}^d$. Let ${\cal C}$ denote the phase space of the classical model, the quantisation of which gives the quantum system under consideration. For our purposes the precise nature of this classical model on ${\cal C}$ is immaterial. Now unitary transformations on Hilbert space are the quantum counterpart of canonical transformations on classical phase space ${\cal C}$ \cite{GIACHETTA}. We may thus regard $SU(d)$ as representing classical canonical transformations,  $\mathbb{C}^d$ being the carrier space of this representation.\footnote{We are considering, as  in ref. \cite{THOOFT2}, the simplified case of a finite--dimensional Hilbert space. Without loss of generality we will restrict to those canonical transformations that are represented by unitary matrices with determinant equal to 1.}

Now quantum states are unit rays rather than vectors, so in fact the true space of inequivalent quantum states is the complex projective space $\mathbb{CP}^{d-1}$.  The latter can be regarded as a homogeneous manifold: 
\begin{equation}
\mathbb{CP}^{d-1}=\frac{SU(d)}{SU(d-1)\times U(1)}.
\label{homogen}
\end{equation}
In this picture we have $SU(d)$ as the total space of a fibre bundle with typical fibre $SU(d-1)\times U(1)$ over the base manifold $\mathbb{CP}^{d-1}$. The projection map 
\begin{equation}
\pi:SU(d)\longrightarrow\mathbb{CP}^{d-1}, \qquad \pi(w):=[w]
\label{projektion}
\end{equation}
arranges points $w\in SU(d)$ into $SU(d-1)\times U(1)$ equivalence classes $[w]$. 

Classical canonical transformations as represented by $SU(d)$ act on the Hilbert space $\mathbb{C}^d$. This descends to an action $\alpha$ of $SU(d)$ 
on $\mathbb{CP}^{d-1}$ as follows:
\begin{equation}
\alpha:SU(d)\times\mathbb{CP}^{d-1}\longrightarrow\mathbb{CP}^{d-1},\quad \alpha\left(u,[v]\right):=[uv].
\label{yeste}
\end{equation}
Here we have $u\in SU(d)$, $[v]\in \mathbb{CP}^{d-1}$, and $uv$ denotes $d\times d$ matrix multiplication.
One readily checks that this action is well defined on the equivalence classes under right multiplication by elements of the stabiliser subgroup $SU(d-1)\times U(1)$. 
This allows one to regard quantum states as equivalence classes of classical canonical transformations on ${\cal C}$. Physically, $u$ in (\ref{yeste}) denotes (the representative matrix of) a canonical transformation on ${\cal C}$, and $[v]$ denotes the equivalence class of (representative matrices of) the canonical transformation $v$ or, equivalently, the quantum state $\vert v\rangle$.  

In the picture just sketched, two canonical transformations are equivalent whenever they differ by a canonical transformation belonging to $SU(d-1)$, and/or whenever they differ by a $U(1)$--transformation. Modding out by $U(1)$ has a clear physical meaning: it is the standard freedom in the choice of the phase of the wavefunction corresponding to the matrix $v\in SU(d)$. Modding out by $SU(d-1)$ also has a physical meaning: canonical transformations on the $(d-1)$--dimensional subspace  $\mathbb{C}^{d-1}\subset\mathbb{C}^d$ are a symmetry of $v$. Therefore the true quantum state $\vert v\rangle$ is obtained from $v\in SU(d)$ after modding out by the stabiliser subgroup $SU(d-1)\times U(1)$.

We conclude that this picture contains some of the elements identified as responsible for the passage from a classical world (canonical transformations) to a quantum world (equivalence classes of canonical transformations, or unit rays within Hilbert space). This is so because some kind of  dissipative mechanism is at work, through the emergence of orbits, or equivalence classes. However the projection (\ref{projektion}) is an on/off mechanism. Instead, one would like to see dissipation occurring as a flow along some continuous parameter.\footnote{None of the above contradicts the fact that the projection (\ref{projektion}) is a smooth map between two differentiable manifolds.} To this end we need some deterministic flow governed by some differential equation. 

We claim that we can render the projection (\ref{projektion}) a dissipative mechanism governed by some differential equation. This equation will turn out to be the Ricci flow (\ref{guillaume}). Proof of this statement follows.

The Lie group $SU(d-1)\times U(1)$ is compact, but it is not semisimple due to the Abelian factor $U(1)$. Leaving the $U(1)$ factor momentarily aside, $SU(d-1)$ is semisimple and compact. As such it qualifies as an Einstein space with positive scalar curvature with respect to the Killing--Cartan metric \cite{BESSE, HELGASON}. Now eqn. (\ref{guillaume}) ensures that $SU(d-1)$ contracts to a point under the Ricci flow.

However  the $U(1)$ factor renders $SU(d-1)\times U(1)$ nonsemisimple. As a consequence, the Killing--Cartan metric of $SU(d-1)\times U(1)$ has a vanishing determinant \cite{HELGASON}. The Ricci flow can still cancel the $SU(d-1)$--factor within $SU(d)$, but not the $U(1)$ factor. After contracting $SU(d-1)$ to a point we are left with the space $U(1)\times \mathbb{CP}^{d-1}$ or, more generally, with a $U(1)$--bundle over the base manifold $\mathbb{CP}^{d-1}$. This $U(1)$--bundle over $\mathbb{CP}^{d-1}$ is the Hopf bundle, where the total space is the sphere $S^{2d-1}$ in $2d-1$ real dimensions \cite{KOBAYASHI}. This sphere falls short of being the true space of quantum states by the unwanted $U(1)$--fibre, that cannot be removed by the Ricci flow. It can, however, be done away with by projection $P$ from the total space of the bundle down to its base. The combined operation ``Ricci flow $R$, followed by projection $P$" acts on the stabiliser subgroup $SU(d-1)\times U(1)$ of the initial $SU(d)$ and leaves us with $\mathbb{CP}^{d-1}$ as desired. Therefore this combined operation $PR$ acts in the same way as the projection $\pi$ in (\ref{projektion}). As opposed to the latter, however, this combined operation $PR$ provides us with a  differential equation that implements dissipation along a continuous parameter, at least along most of the way. 

To conclude we would like to make some topological remarks. One could expect the projection $P$ to be necessary since the fundamental group of $U(1)$ is the group of integers $\mathbb{Z}$. The topological charge carried by the latter cannot be done away with continuously. One the other hand, any continuous deformation of $SU(d-1)$ to a point\footnote{Or possibly to a finite number of points}  would also have done the job. However, the Ricci flow $R$ is particularly interesting because, as we will see presently, we can identify the Ricci tensor as its infinitesimal generator, or Hamiltonian.

\section{Discussion}\label{unterhaltung}

Our starting point was the observation that canonical transformations on classical phase space are implemented quantum--mechanically as unitary transformations on the Hilbert space of quantum states. In our finite--dimensional setup,  this gave rise to a natural action of $SU(d)$ on $\mathbb{C}^d$. This action  provided us with the building blocks to construct the deterministic system that we take to underlie the given quantum mechanics. Next, different pieces of classical information (elements of $SU(d)$, or classical canonical transformations) were arranged into quantum equivalence classes (points on $\mathbb{CP}^d$, or quantum states): this procedure implements information loss, or dissipation. Quantum states thus arose as equivalence classes of canonical transformations on classical phase space. However, dissipation was not implemented by means of the usual projection (\ref{projektion}) (an on/off mechanism), but rather by means of the Ricci flow (followed by the projection $P$). The rationale was that the Ricci flow provided us with a a deterministic mechanism governed by a dissipative differential equation, that can be understood as a flow along a continuous parameter. 

In a nutshell, our deterministic model is the group manifold $SU(d)$, acted on by the combined operation $PR$ described above. Here $R$ stands for the Ricci flow of $SU(d-1)$ down to one point, and $P$ stands for the projection from the Hopf bundle with total space $S^{2d-1}$ onto its base $\mathbb{CP}^{d-1}$.  In our setup, the Ricci tensor acts as a Hamiltonian for the Ricci flow. Since contraction to a point requires a positive definite Ricci tensor, also the Hamiltonian is positive definite. That the Hamiltonian is bounded from below is an immediate consequence of the finite dimensionality $d$ of the given Hilbert space. However, were one to take the limit $d\to\infty$, boundedness from below would still be guaranteed by the fact that the Ricci tensor is assumed positive definite. Thus positive curvature guarantees the existence of a stable ground state.

There is an added bonus to having positive curvature. Namely, it is a geometrical property that we do not have to impose, since the special unitary group $SU(d)$
comes naturally endowed with a metric that ensures it. The same can be said of the complex projective space $\mathbb{CP}^{d-1}$. It is amusing to ponder to what extent the requirement of positive curvature imposes one particular set of canonical transformations with its corresponding space of quantum states (namely, the special unitary group and projective space)---or rather the opposite (namely, since canonical transformations are unitary matrices and quantum states are unit rays, positive curvature follows naturally). Geometrically this question is of no import, and one can take either view as the starting point. However, for the quantum theory, which starting point one takes is decisive. The point of view adopted here is that curvature comes first and quantum states follow, since this order of things ensures the correct arrangement of classical objects into quantum equivalence classes. It is thus no coincidence that a positively--curved Einstein manifold, complex projective space, is the true space of inequivalent quantum states. 

Our previous considerations apply to the space of quantum states, regardless of spacetime. However, curvature being primary rather than secondary, one can wonder what consequences this may have on a would--be quantum theory of {\it spacetime}\/. Are we supposed to {\it quantise}\/ a classical theory of gravity? Should we not {\it relativise}\/ the notion of a quantum instead?  Bohr's principle of complementarity suggests that both these two viewpoints (a quantum theory of gravity {\it vs.}\/ a gravitational theory of quanta), though mutually excluding, should be taken into account. Our analysis of ref.  \cite{ME} is a contribution to this latter viewpoint, which the conclusions of this paper also support.

\vskip.5cm
\noindent
{\bf Acknowledgements} J.M.I is pleased to thank Max--Planck--Institut f\"ur Gravitationsphysik, Albert--Einstein--Institut (Potsdam, Germany) for hospitality extended over a long period of time.

\end{document}